\documentstyle[aps,12pt,amssymb,epsfig]{revtex}

\begin{document}

\baselineskip=0.75cm
\hfill DSF-22/2002
\vspace{0.5 cm}

\centerline{{\bf PENTAGON INTEGRALS FOR HEAVY QUARK PHYSICS}}

\vspace{1 cm}

\centerline{\large{Francesco Tramontano$^*$}}

\footnotetext[1]{e-mail: tramontano@na.infn.it}

\vspace{0.5 cm}

\centerline{Dipartimento di Scienze Fisiche, Universit\`{a} di Napoli,}
\centerline{Complesso di Monte S. Angelo, Via Cintia, Napoli, Italy}

\vspace{2 cm}

\begin{abstract}
\begin{center} 
{\bf Abstract}
\end{center}

\noindent
In this paper we present the calculation of a scalar pentagon integral
with two consecutive massive external legs having an equal mass propagator
embedded between them. We also deal with the two situations where the
farest external leg is either massive or not. The relevance of the
calculation comes from its application in many perturbative QCD
calculations as well as in QCD corrections for weak precesses.

\end{abstract}


\newpage

\section{Introduction}

A series of rare elementary processes involving more than
two particles in the final state are going to be measured with
increasing precision. The multiplicity of the final state makes it
difficult to extract predictions by the standard gauge theories
even if semplifications arise when either partecipants are all massless
or only some of the external particles are massive. However more accurate
rate measurements of processes with heavy quark hadrons in the final state
will soon be available as is the case of the CHORUS experiment where
direct evidence for the associate charm production in charged
current neutrino nucleon scattering has been shown \cite{CHO}. In the one
loop calculation of such processes we encounter pentagon integrals with a
massive line as skeched in figure 1 where massive particles are bold,
massless ones thin and dashed ones can be either massive or not.
\begin{figure}[here]
\begin{center}
\begin{minipage}{6.5cm}
\includegraphics[width=6cm]{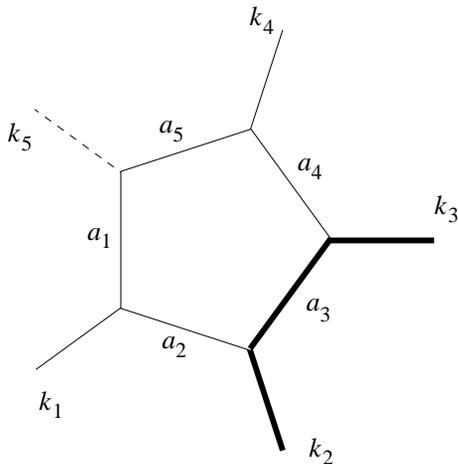}
\end{minipage}
\caption{Pentagon with one massive line}
\end{center}
\end{figure}

In general the inclusion of masses makes things more involved, although
the calculation simplifies when either external masses are equal to each
other or they are equal to the internal masses or both eventualities occur
as it is often the case in normal gauge theories. Recently a lot of
progress has been made in the technics for perturbative calculations with
different approaches. A non-comprehensive list is given in \cite{MP} and
reference therein and in \cite{BDK}, \cite{POL}, \cite{DAV}, \cite{ND}
and \cite{PAS}. In
particular adopting the dimensional regularization approach for Feynman
parametrized integrand the authors of ref.\cite{BDK} derived
simplifications and recursion formulas by the implementation of algebraic
technic. Using these methods the problem of the evaluation of a one loop 
$n$ points scalar integral is translated to the evaluation of a
combination of
$n-1$ points scalar integrals and the original $n$ points integral in
$D=6-2\,\varepsilon$ dimensions; moreover the original $n$ points
one loop integral
can be represented as the solution of a partial differential equation
system. In the present paper we use this approach to perform the
calculation of the pentagon integral represented in figure 1. Other
massive pentagon integrals have been recently evaluated in next to
leading order calculations of processes in which an Higgs particle
can be generated at hadron colliders. In particular two
independent groups report the NLO corrections for the process in which an
Higgs particle is generated together with a $t~\overline{t}$ pair, \cite{RDWO}
and \cite{BDKPSZ}. Another NLO calculation involving massive pentagon
integrals is given in \cite{DKOSZ} in which the final state considered
consists of an Higgs particle plus two jets.
The general methods employed here do not concern with the specific processes
and  the results must be analitically continued to describe a specific
process. Finally only the most simple tensor integral is given
while we postpone other cases to a dedicated paper \cite{INP}. The paper
is organized as follow: in section II relevant formulas from
ref.\cite{BDK} are collected, in section III they are applied to the
scalar massive pentagon represented in figure 1 transforming it in a
combination of four points integrals; section IV is devoted to
four point integrals evaluation and in section V more simple tensor
integral (vector) is given with the conclusions. The initial condition
for the differential equations originated in the four points
evaluation are calculated in the appendix.

\section{Basic formulas}

The starting point is the integral in $D=4-2\varepsilon$ dimensions
\begin{equation} \label{iniz}
  {\mathcal I}_n=\mu^{2\varepsilon} \int
  \frac{d^{4-2\varepsilon}l}{(2\pi)^{4-2\varepsilon}}
  \,\frac{1}{(l^2-M_1^2)((l-p_1)^2-M_2^2)...((l-p_{n-1})^2-M_n^2)}
\end{equation}
with the momenta $k_i$ taken to be outgoing, $k_i^2=m_i^2$ and
\begin{eqnarray}\label{pi}
  p_i \equiv \sum_{j=1}^{i}k_j \nonumber \\
  p_0=p_n=0.
\end{eqnarray}
Applying Feynman parametrization, Wick rotating and
integrating over loop momentum this integral can be cast in the form
\begin{equation}
  {\mathcal I}_n =
  i\,(-1)^n\,(4\pi)^{\varepsilon-2}\,\mu^{2\varepsilon}\,I_n
\end{equation}
\begin{equation} \label{quat}
  I_n=\Gamma(n-2+\varepsilon)\int_0^1 d^na_i \,\delta(1-\sum_ia_i)
  \,\frac{1}{{\mathcal D}(a_i)^{n-2+\varepsilon}} \label{aaa}
\end{equation}
with
\begin{equation}
  {\mathcal D}(a_i)=\sum_{i,j=1}^nS_{ij}\,a_i\,a_j \label{ccc}
\end{equation}
and the matrix $S$ given by
\begin{equation}
  S=\frac{1}{2}\,(M_i^2+M_j^2-p_{ij}^2)
\end{equation}
with $p_{ii}=0$ and
\begin{equation}
  p_{ij}\equiv k_i+k_{i+1}+...+k_{j-1}
\end{equation}
for $i<j$. We will not repeat the derivations obtained in
ref.\cite{BDK} but, to introduce notation and to be self-consistent, in
the rest of this section we just collect relevant formulas that will be
used in section III and IV. Performing a projective transformation
\cite{tHV} with parameters $\alpha_i$ in such a way that the
denominator in Eq.(\ref{aaa}) has no $\alpha_i$ dependence the
definition of a new matrix follows (indices are not summed)
\begin{equation} \label{rrhhoo}
  \rho_{ij}=S_{ij}\,\alpha_i\,\alpha_j.
\end{equation}
Using the following definitions
\begin{eqnarray}\label{fff}
  \hat{I}_n=\left( \prod_{j=1}^n\alpha_j \right)^{-1}\,I_n\nonumber\\
  N_n=2^{n-1} \, det(\rho)\nonumber\\
  \eta=N_n \,\rho^{-1}\\
  \gamma_i=\sum_{j=1}^n\eta_{ij}\,\alpha_j\nonumber\\
  \hat{\Delta}_n=\sum_{i,j=1}^n\eta_{ij}\,\alpha_i\,\alpha_j\nonumber
\end{eqnarray}
the authors of ref.\cite{BDK} find
\begin{eqnarray}
  \hat{I}_n=\frac{1}{2N_n}\left[\sum_{i=1}^n\gamma_i\,\hat{I}_{n-1}^{(i)}+
  (n-5+2\varepsilon)\,\hat{\Delta}_n\,\hat{I}_n^{D=6-2\varepsilon}\right]\\
  \frac{1}{n-4+2\varepsilon}\,\frac{\partial\hat{I}_n}{\partial\alpha_i}=
  \frac{1}{2N_n}\left[\sum_{j=1}^n\eta_{ij}\,\hat{I}_{n-1}^{(j)}+
  (n-5+2\varepsilon)\,\gamma_i\,\hat{I}_n^{D=6-2\varepsilon}\right]
\end{eqnarray}
where $\hat{I}_{n-1}^{(i)}$ stands for the $n-1$ integral with the
denominator obtained from an $\hat{I}_{n}$ integral eliminating the
propagator between legs $i-1$ and $i$; once Feynman parameter has
been introduced in the usual way for $\hat{I}_{n}$ the denominator in
$\hat{I}_{n-1}^{(i)}$ is obtained putting $a_i=0$. By the observation that
$\hat{I}_4$ and $\hat{I}_5$ are finite in $6$ dimensions, performing
one-loop calculation one can limit to evaluate
\begin{eqnarray}
  \hat{I}_5=\frac{1}{2N_5}\sum_{i=1}^5\gamma_i\,\hat{I}_{4}^{(i)}+
  {\mathcal O}(\varepsilon) \label{ppp}\\
  \frac{\partial\hat{I}_4}{\partial\alpha_i}=
  \frac{\varepsilon}{N_4}\sum_{j=1}^4\eta_{ij}\,\hat{I}_{3}^{(j)}+
  {\mathcal O}(\varepsilon)\label{bbb}
\end{eqnarray}
taking only the divergent part from the $\hat{I}_{3}^{(j)}$
integrals in Eq.(\ref{bbb}).

\section{Pentagon with one massive line}

To write down the integral in figure 1 we set
$k_1^2=k_4^2=M_1^2=M_2^2=M_4^2=M_5^2=0$,
$k_2^2=k_3^2=M_3^2=m^2$ and $k_5^2=q^2$ giving
\begin{equation}
  {\mathcal I}_5=\mu^{2\varepsilon} \int
  \frac{d^{4-2\varepsilon}l}{(2\pi)^{4-2\varepsilon}}
  \,\frac{1}{(l^2)(l-p_1)^2((l-p_2)^2-m^2)(l-p_3)^2(l-p_4)^2}
\end{equation}
and
\begin{equation}
  I_5=\Gamma(3+\varepsilon)\int_0^1 d^5a_i \,\delta(1-\sum_ia_i)
  \,\frac{1}{{\mathcal D}(a_i)^{3+\varepsilon}}
\end{equation}
with ${\mathcal D}$ given in Eq.(\ref{ccc}) and the matrix $S$ given by
\begin{equation}
  S=\frac{1}{2}\pmatrix{0&0&m^2-s_{12}&-s_{45}&-q^2\cr 0&0&0&-s_{23}&-s_{51}
  \cr m^2-s_{12}&0&2m^2&0&m^2-s_{34}\cr -s_{45}&-s_{23}&0&0&0
  \cr -q^2&-s_{51}&m^2-s_{34}&0&0}\label{ddd}
\end{equation}
with $s_{i,i+1}=(k_i+k_{i+1})^2$.
\noindent
We define $\bar{s}_{i,i+1}=s_{i,i+1}-m^2$ and in the following we will
assume $\bar{s}_{12},s_{23},\bar{s}_{34},s_{45},s_{51},q^2<0$.
Performing the projective transformation with
\begin{eqnarray}\label{c1}
  \alpha_1=\sqrt{-\frac{s_{23}\,\bar{s}_{34}}{s_{45}\,s_{51}\,\bar{s}_{12}}}\nonumber\\
  \alpha_2=\sqrt{-\frac{s_{45}\,\bar{s}_{34}}{s_{23}\,s_{51}\,\bar{s}_{12}}}\nonumber\\
  \alpha_3=\sqrt{-\frac{s_{45}\,s_{51}}{s_{23}\,\bar{s}_{34}\,\bar{s}_{12}}}\\
  \alpha_4=\sqrt{-\frac{s_{51}\,\bar{s}_{12}}{s_{23}\,\bar{s}_{34}\,s_{45}}}\nonumber\\
  \alpha_5=\sqrt{-\frac{\bar{s}_{12}\,s_{23}}{\bar{s}_{34}\,s_{45}\,s_{51}}}\nonumber
\end{eqnarray}
we get for the $\rho_{ij}$ matrix in Eq.(\ref{rrhhoo})
\begin{equation}
  \rho=\frac{1}{2}\pmatrix{0&0&1&1&\lambda\cr 0&0&0&1&1
  \cr 1&0&2M^2&0&1\cr 1&1&0&0&0
  \cr \lambda&1&1&0&0}\label{eee}
\end{equation}
with
\begin{equation}\label{c2}
  \lambda=\frac{q^2\,s_{23}}{s_{45}\,s_{51}}
\end{equation}
\begin{equation}\label{c3}
  M^2=-\,\frac{m^2\,s_{45}\,s_{51}}{\bar{s}_{12}\,s_{23}\,\bar{s}_{34}}
\end{equation}
If $k_5^2=q^2=0$ we only have to take $\lambda=0$ in
Eq.(\ref{eee}). The coefficient relevant for the evaluation of the
pentagon by Eq.(\ref{ppp}) are given in the table 1, keeping apart the
case $q^2=0$.
Due to the presence of masses we have not cyclic relations between the
coefficients but only the relations
\begin{eqnarray}
  \gamma_4= \left. \gamma_2 \right|_{{\footnotesize  
  \alpha_4\leftrightarrow \alpha_2,~ \alpha_5 \leftrightarrow \alpha_1 }}
  \nonumber \\
  \gamma_5= \left. \gamma_1 \right|_{{\footnotesize
  \alpha_4\leftrightarrow \alpha_2,~ \alpha_5 \leftrightarrow \alpha_1 }}
\end{eqnarray}
{\small
\begin{center}
Table 1. Coefficients to be used in Eq.(\ref{ppp})
\begin{tabular}{|c|c|c|}
\hline
  par & any $q^2$ & $q^2=0$ \\
      &   &   \\ \hline
  $N_5$ & $1 + M^2 - \left( 1 + 2\,M^2 \right) \,\lambda  + M^2\,{\lambda 
  }^2$ & $1 + M^2$ \\ &   &   \\ \hline
  $\gamma_1$ & $\begin{array}{c}{{\alpha }_1} - {{\alpha }_2} + {{\alpha }_3} -
  \lambda \,{{\alpha }_3} +
  {{\alpha }_4} + 2\,M^2\,{{\alpha }_4} \\ - 2\,M^2\,\lambda \,{{\alpha }_4} -
  {{\alpha }_5} - 2\,M^2\,{{\alpha }_5} + 2\,M^2\,\lambda \,{{\alpha
  }_5}\end{array}$ & ${{\alpha }_1}
  - {{\alpha }_2} + {{\alpha }_3} +
  \left( 1 + 2\,M^2 \right) \,\left( {{\alpha }_4} - {{\alpha }_5} \right)$ \\ \hline
  $\gamma_2$ & $\begin{array}{c}-{{\alpha }_1} + {{\alpha }_2} - {{\alpha }_3} + \lambda \,{{\alpha }_3} +
  {{\alpha }_4} - 2\,\lambda \,{{\alpha }_4} - 2\,M^2\,\lambda \,{{\alpha }_4} \\ +
  2\,M^2\,{\lambda }^2\,{{\alpha }_4} + {{\alpha }_5} + 2\,M^2\,{{\alpha }_5} -
  2\,M^2\,\lambda \,{{\alpha }_5}\end{array}$ & $-{{\alpha }_1} + {{\alpha }_2} - {{\alpha }_3} +
  {{\alpha }_4} + (1 + 2\,M^2\,){{\alpha }_5}$ \\ \hline
  $\gamma_3$ & $\begin{array}{c}{{\alpha }_1} - \lambda \,{{\alpha }_1} - {{\alpha }_2}
  + \lambda \,{{\alpha }_2} +
  {{\alpha }_3} - 2\,\lambda \,{{\alpha }_3} \\+ {\lambda }^2\,{{\alpha }_3} -
  {{\alpha }_4} + \lambda \,{{\alpha }_4} + {{\alpha }_5}
  - \lambda \,{{\alpha }_5}\end{array}$ & ${{\alpha }_1}
  - {{\alpha }_2} + {{\alpha }_3} - {{\alpha }_4} + {{\alpha }_5}$ \\ \hline
  $\gamma_4$ & $\begin{array}{c}{{\alpha }_1} + 2\,M^2\,{{\alpha }_1} - 2\,M^2\,\lambda \,{{\alpha }_1} +
  {{\alpha }_2} - 2\,\lambda \,{{\alpha }_2} \\- 2\,M^2\,\lambda \,{{\alpha }_2} +
  2\,M^2\,{\lambda }^2\,{{\alpha }_2} - {{\alpha }_3} + \lambda \,{{\alpha }_3} +
  {{\alpha }_4} - {{\alpha }_5}\end{array}$ & $\left( 1 + 2\,M^2 \right) \,{{\alpha }_1} + {{\alpha }_2}
  - {{\alpha }_3} + {{\alpha }_4} - {{\alpha }_5}$ \\ \hline
  $\gamma_5$ & $\begin{array}{c}-{{\alpha }_1} - 2\,M^2\,{{\alpha }_1} + 2\,M^2\,\lambda \,{{\alpha }_1} +
  {{\alpha }_2} + 2\,M^2\,{{\alpha }_2} \\ - 2\,M^2\,\lambda \,{{\alpha }_2}+
  {{\alpha }_3} - \lambda \,{{\alpha }_3} - {{\alpha }_4} + {{\alpha }_5}\end{array}$ &
  $\left( 1 + 2\,M^2 \right) (\,{{\alpha }_2} - \,{{\alpha }_1} )+ {{\alpha }_3} - {{\alpha }_4} +
  {{\alpha }_5}$ \\ \hline
\end{tabular}
\end{center}}

\vspace{0.5cm}

\noindent
In terms of new kinematical variables $\alpha_i$, $\lambda$ and $M$
the denominator in the $\hat{I}_5$ integral represented in Eq.(\ref{ccc})
is given by
\begin{equation}\label{ggg}
  \frac{M^2\,{{a_3}}^2}{{{{\alpha }_3}}^2} +
  \frac{{a_1}\,{a_3}}{{{\alpha }_1}\,{{\alpha }_3}} +
  \frac{{a_1}\,{a_4}}{{{\alpha }_1}\,{{\alpha }_4}} +
  \frac{{a_2}\,{a_4}}{{{\alpha }_2}\,{{\alpha }_4}} +
  \frac{\lambda \,{a_1}\,{a_5}}
   {{{\alpha }_1}\,{{\alpha }_5}} +
  \frac{{a_2}\,{a_5}}{{{\alpha }_2}\,{{\alpha }_5}} +
  \frac{{a_3}\,{a_5}}{{{\alpha }_3}\,{{\alpha }_5}}
\end{equation}
and the four points denominators in the $\hat{I}_4^{(i)}$ integrals in
Eq.(\ref{ppp}) can be obtained putting $a_i$ to zero in the expression
above. It is easy to verify the relations
\begin{eqnarray}\label{rel}
  \hat{I}_4^{(4)}(\alpha_1,\alpha_2,\alpha_3,\alpha_5)=
  \left. \hat{I}_4^{(2)}(\alpha_1,\alpha_3,\alpha_4,\alpha_5)
  \right|_{{\footnotesize
  \alpha_4\rightarrow \alpha_2,~ \alpha_5
  \leftrightarrow \alpha_1 }} \nonumber \\
  \hat{I}_4^{(5)}(\alpha_1,\alpha_2,\alpha_3,\alpha_4)=
  \left. \hat{I}_4^{(1)}(\alpha_2,\alpha_3,\alpha_4,\alpha_5)
  \right|_{{\footnotesize
  \alpha_5\rightarrow \alpha_1,~ \alpha_4
  \leftrightarrow \alpha_2 }}.
\end{eqnarray}
In the next section we proceed to the evaluation of $\hat{I}_4^{(1)}$,
$\hat{I}_4^{(2)}$ and $\hat{I}_4^{(3)}$ using the set of partial
differential Eqs.(\ref{bbb}).

\section{Four points functions}

Here we evaluate the integrals $\hat{I}_4^{(1)}$ and $\hat{I}_4^{(2)}$,
corrsponding to
massive boxes with an internal massive line, in the
variables defined in Eqs.(\ref{c1}, \ref{c2}, and \ref{c3}) and 
translate the integrals $\hat{I}_4^{(3)}$ that are well known and
correspond to massive
boxes with massles internal lines.

\subsection{The integral $\hat{I}_4^{(1)}$}

After putting $a_1=0$ in Eq.(\ref{ggg}) we have for the denominator
in $\hat{I}_4^{(1)}$
\begin{equation}
  \frac{M^2\,{{a_3}}^2}{{{{\alpha }_3}}^2} +
  \frac{{a_2}\,{a_4}}{{{\alpha }_2}\,{{\alpha }_4}} +
  \frac{{a_2}\,{a_5}}{{{\alpha }_2}\,{{\alpha }_5}} +
  \frac{{a_3}\,{a_5}}{{{\alpha }_3}\,{{\alpha }_5}}.
\end{equation}
Before solving the integral we perform the following kinematic transformation:
\begin{eqnarray} \label{tr1}
  \alpha_2&=&M\,c_2 \nonumber \\
  \alpha_3&=&M\,c_3 \nonumber \\
  \alpha_4&=&c_4/M \nonumber \\
  \alpha_5&=&c_5/M.\nonumber \\
\end{eqnarray}
In terms of the new variables we get for the denominator:
\begin{equation}
  \frac{{{a_3}}^2}{{{{c}_3}}^2} +
  \frac{{a_2}\,{a_4}}{{{c}_2}\,{{c}_4}} +
  \frac{{a_2}\,{a_5}}{{{c}_2}\,{{c}_5}} +
  \frac{{a_3}\,{a_5}}{{{c}_3}\,{{c}_5}}.
\end{equation}
and
\begin{eqnarray}
  N_4&=&1/2 \nonumber \\
  \eta&=& \pmatrix{0&0&1&0\cr 0&0&-1&1\cr 1&-1&-2&2\cr
0&1&2&-2}.
\end{eqnarray}
The only divergent three points functions extracted by the expression
above are $\hat{I}_3^{(1)}$ and $\hat{I}_3^{(2)}$ obtained putting $a_2=0$
and $a_3=0$ respectively;
these correspond to two two-mass triangles, while the other two
obtained putting $a_4=0$ and $a_5=0$ respectively are three-mass
triangles checked to be finite. At the ${\mathcal O}(\varepsilon^{-1})$
we have
\begin{eqnarray} \label{in1}
  \hat{I}_3^{(1)}&=&\Gamma(1+\varepsilon)\left(
  \frac{1}{2 \, \varepsilon^2 c_4}+
  \frac{\log(c_5)}{\varepsilon \, c_4}
  \right) + {\mathcal O}(\varepsilon^0)\\
  \hat{I}_3^{(2)}&=&
  \frac{\Gamma(1+\varepsilon)}{\varepsilon(c_4-c_5)}
  \log \left( \frac{c_5}{c_4} \right)
  + {\mathcal O}(\varepsilon^0).
\end{eqnarray}
The system of partial differential equations in Eq.(\ref{bbb}) is then
given by
\begin{eqnarray}
  \frac{\partial\hat{I}_4^{(1)}}{\partial c_2}&=&0 \nonumber \\
  \frac{\partial\hat{I}_4^{(1)}}{\partial c_3}&=&0 \nonumber \\
  \frac{\partial\hat{I}_4^{(1)}}{\partial c_4}&=
  &\frac{\Gamma(1+\varepsilon)}{c_4-c_5}\left[
  \frac{1}{\varepsilon}\left( 1-\frac{c_5}{c_4} \right)+
  2\log(c_4)-2 \, \frac{c_5}{c_4}\log(c_5)
  \right] \nonumber \\
  \frac{\partial\hat{I}_4^{(1)}}{\partial c_5}&=&
  \frac{2\,\Gamma(1+\varepsilon)}{c_4-c_5}
  \, \log \left( \frac{c_5}{c_4} \right)
\end{eqnarray}
with the solution
\begin{eqnarray}\label{sol1}
  \hat{I}_4^{(1)} &=&\Gamma(1+\varepsilon)\left[
  \frac{1}{\varepsilon}\log(c_4)+
  \left( \log\left(1-\frac{c_4}{c
  _5}\right)-\log(c_4-c_5)\right)^2 +
  2\log(c_4)\log(c_5-c_4) +
  \right. \nonumber \\
  && \left. 2\,\textup{Li}_2\left(\frac{c_4}{c_5}\right)-
  2\log(c_5)\log(c_4-c_5) + k_1 \right].
\end{eqnarray}
The integration constant $k_1$ is evaluated in the appendix and its
value is
\begin{equation}\label{c01}
  k_{1}=\frac{1}{2\, \varepsilon^2} + 5\,\zeta(2)
\end{equation}
where $\textup{Li}_2$ is the dilogarithm function and $\zeta(2)=\pi^2/6$.
Afetr some manipulation we have
\begin{equation}
\Gamma(1 + \varepsilon)\,\left(
    \frac{{{c_4}}^{2\,\varepsilon}}{2\,\varepsilon^2} - 
    {\log^2 \left(\frac{{c_4}}{{c_5}}\right)} - 
    2\,\textup{Li}_2 \left( 1 - \frac{{c_4}}{{c_5}} \right) +
    \frac{{\pi }^2}{6}
    \right)
\end{equation}
Reintroducing the original variables inverting Eq.(\ref{tr1})
we get for $\hat{I}_4^{(1)}$:
\begin{equation}
\hat{I}_4^{(1)}=\Gamma(1 + \varepsilon)\,\left(
    \frac{{{(M\,\alpha_4)}}^{2\,\varepsilon}}{2\,\varepsilon^2} - 
    {\log^2 \left(\frac{{\alpha_4}}{{\alpha_5}}\right)} - 
    2\,\textup{Li}_2 \left( 1 - \frac{{\alpha_4}}{{\alpha_5}} \right)
    + \frac{{\pi }^2}{6}
    \right)
\end{equation}
Being $\hat{I}_4^{(1)}$ independent from $\lambda$ its value does not
change in the limit $q^2\rightarrow 0$.

\subsection{The integral $\hat{I}_4^{(2)}$}

Here and in the following subsection we proceed performing the same
steps as in the derivation of $\hat{I}_4^{(1)}$. The limit
$q^2\rightarrow 0$ now gives a different situation; in fact in this
limit there will be three divergent three-point integrals extracted by
$\hat{I}_4^{(2)}$ so as explained in \cite{BDK} the limit procedure is
not smooth and the two case have to be taken separately.

\subsubsection{$\hat{I}_4^{(2)}$, $q^2\neq 0$}

In this case $\hat{I}_4^{(2)}$ is a three external mass box but,
differently from $\hat{I}_4^{(1)}$, it has all external masses
different from each other and so it needs evaluation.
After putting $a_2=0$ in Eq.(\ref{ggg}) we have for the denominator in
$\hat{I}_4^{(2)}$
\begin{equation}\label{den2}
  \frac{M^2\,{{a_3}}^2}{{{{\alpha }_3}}^2} +
  \frac{{a_1}\,{a_3}}{{{\alpha }_1}\,{{\alpha }_3}} +
  \frac{{a_1}\,{a_4}}{{{\alpha }_1}\,{{\alpha }_4}} +
  \frac{\lambda \,{a_1}\,{a_5}}{{{\alpha }_1}\,{{\alpha }_5}} +
  \frac{{a_3}\,{a_5}}{{{\alpha }_3}\,{{\alpha }_5}}.
\end{equation}
Rescaling the variables with
\begin{eqnarray} \label{tr2}
  \alpha_1&=&c_1/M \nonumber \\
  \alpha_3&=&M\,c_3 \nonumber \\
  \alpha_4&=&M\,c_4 \nonumber \\
  \alpha_5&=&c_5/M \nonumber \\
  \lambda&=&\delta/M^2
\end{eqnarray}
we get
\begin{equation}
  \frac{{{a_3}}^2}{{{{c}_3}}^2} +
  \frac{{a_1}\,{a_3}}{{{c}_1}\,{{c}_3}} +
  \frac{{a_1}\,{a_4}}{{{c}_1}\,{{c}_4}} +
  \frac{\delta\,{a_1}\,{a_5}}{{{c}_1}\,{{c}_5}} +
  \frac{{a_3}\,{a_5}}{{{c}_3}\,{{c}_5}}.
\end{equation}
giving $N_4=1/2$ and
\begin{equation}\label{et2}
  \eta = \pmatrix{0&0&1&0\cr 0&0&-\delta&1\cr
  1&-\delta&2\delta(1- \delta)&2\delta -1\cr
  0&1&2\delta-1&-2}.
\end{equation}
The only divergent three points functions extracted by the expression
above are $\hat{I}_3^{(1)}$ and
$\hat{I}_3^{(2)}$ obtained putting $a_1=0$ and $a_3=0$ respectively;
these correspond to two two-mass triangle, while the other two
obtained putting $a_4=0$ and $a_5=0$ respectively are three-mass
triangle checked to be finite. At the ${\mathcal O}(\varepsilon^{-1})$
we have
\begin{eqnarray}
  \hat{I}_3^{(1)}&=&\Gamma(1+\varepsilon)\left(
  \frac{1}{2 \, \varepsilon^2 c_4}+
  \frac{\log(c_5)}{\varepsilon \, c_4}
  \right) + {\mathcal O}(\varepsilon^0)\\
  \hat{I}_3^{(2)}&=&\frac{\Gamma(1+\varepsilon)}{\varepsilon
  (c_5-\delta c_4)}
  \log\left(\frac{\delta c_4}{c_5}
  \right) + {\mathcal O}(\varepsilon^0).\\
\end{eqnarray}
The system in Eq.(\ref{bbb}) is then given by
\begin{eqnarray}
  \frac{\partial\hat{I}_4^{(2)}}{\partial c_1}&=&0 \nonumber \\
  \frac{\partial\hat{I}_4^{(2)}}{\partial c_3}&=&0 \nonumber \\
  \frac{\partial\hat{I}_4^{(2)}}{\partial c_4}&=&
  \frac{\Gamma(1+\varepsilon)}{c_5-\delta \,c_4}\left[
  \frac{1}{\varepsilon}\left( \frac{c_5}{c_4} -\delta  \right)+
  2\,\frac{c_5}{c_4}\log(c_5)-2\delta \log(\delta
  \,c_4) \right] \nonumber \\
  \frac{\partial\hat{I}_4^{(2)}}{\partial c_5}&=&
  \frac{2\,\Gamma(1+\varepsilon)}{c_5-\delta \,c_4}
  \log \left( \frac{\delta \,c_4}{c_5} \right)
\end{eqnarray}
with the solution
\begin{eqnarray} \label{sol2}
  \hat{I}_4^{(2)} &=&\Gamma(1+\varepsilon)\left[
  \frac{1}{\varepsilon}\log(c_4)+ 2\log(c_4)\log(c_5)
  - 2\log(c_5- \delta \,c_4)\log\left( 1- \frac{\delta
  \,c_4}{c_5} \right) \right. \nonumber\\
  && \left. +2\,\textup{Li}_2\left( \frac{\delta \,c_4}{c_5} \right)
  +\log^2(c_5-\delta\,c_4) - 2\,\log(c_5)\,\log(c_5-\delta\,c_4)
  \right. \nonumber\\
  && \left. +\log^2\left( 1- \frac{\delta \,c_4}{c_5} \right) + 
  2\,\log(\delta)\,\log(c_5-\delta\,c_4) +
  2\,\log(c_4)\,\log\left( 1- \frac{\delta \,c_4}{c_5} \right)
  +k_2 \right].
\end{eqnarray}
The integration constant $k_{2}$ is evaluated in the appendix and its
value is
\begin{equation}\label{c02}
  k_{2}=\frac{1}{2\,\varepsilon^2}+
  \frac{1}{\varepsilon}\log(\delta)
  - \frac{1}{2}\log^2(\delta) -\,\textup{Li}_2( 1\,-\,\delta ) - 2\,\zeta(2).
\end{equation}
After some maipulation we have
\begin{equation}
\hat{I}_4^{(2)} =\Gamma(1+\varepsilon) \left[ 
  \left( \frac{1}{\varepsilon^2} - \frac{1}{\varepsilon} + 1 \right)
\,\frac{{\left( {c_4}\,\delta  \right) }^{2\,\varepsilon}}
  {2} - \frac{{\log^2 (\delta )}}{2} - {\log^2 \left(\frac{{c_4}\,\delta }{{c_5}}\right)} - 
  \textup{Li}_2 ( 1 - \delta ) - 2\,
   \textup{Li}_2 \left( 1 - \frac{{c_4}\,\delta }{{c_5}} \right)
  \right]
\end{equation}
Reintroducing the original variables inverting Eq.(\ref{tr2})
we get for $\hat{I}_4^{(2)}$:
\begin{eqnarray}
\hat{I}_4^{(2)}&=&\Gamma(1+\varepsilon) \left[ 
  \left( \frac{1}{\varepsilon^2} - \frac{1}{\varepsilon} +1 \right)
\,\frac{{\left( M\,\lambda\,{\alpha_4} \right) }^{2\,\varepsilon}}
  {2} - \frac{1}{2}\log^2 (M^2\,\lambda ) - {\log^2 \left( \frac{\lambda\, {\alpha_4} }{{\alpha_5}}
  \right) } \right. \nonumber\\
  &&\left. -\,\textup{Li}_2 ( 1 - M^2\, \lambda ) - 2\,
   \textup{Li}_2 \left( 1 - \frac{\lambda\,{\alpha_4} }{{\alpha_5}} \right)
  \right] \\
\end{eqnarray}

\subsubsection{$\hat{I}_4^{(2)}$, $q^2= 0$}

In this case $\hat{I}_4^{(2)}$ is a two external mass box. Putting
$\lambda=0$ the denominator in Eq.(\ref{den2}) became
\begin{equation}
  \frac{M^2\,{{a_3}}^2}{{{{\alpha }_3}}^2} +
  \frac{{a_1}\,{a_3}}{{{\alpha }_1}\,{{\alpha }_3}} +
  \frac{{a_1}\,{a_4}}{{{\alpha }_1}\,{{\alpha }_4}} +
  \frac{{a_3}\,{a_5}}{{{\alpha }_3}\,{{\alpha }_5}}
\end{equation}
Rescaling the variables as in Eq.(\ref{tr2}) we get
\begin{equation}
\frac{{{a_3}}^2}{{{{c}_3}}^2} +
\frac{{a_1}\,{a_3}}{{{c}_1}\,{{c}_3}} +
\frac{{a_1}\,{a_4}}{{{c}_1}\,{{c}_4}} +
\frac{{a_3}\,{a_5}}{{{c}_3}\,{{c}_5}}
\end{equation}
The divergent three point functions $\hat{I}_3^{(1)}$,
$\hat{I}_3^{(2)}$ and $\hat{I}_3^{(3)}$ are obtained putting $a_1$,
$a_2$ and $a_3=0$ respectively.
\begin{eqnarray}
  \hat{I}_3^{(1)}&=&\Gamma(1+\varepsilon)\left(\frac{1}{\varepsilon^2\,c_4}+
  \frac{1}{\varepsilon\,c_4}\log(c_5)\right)
  + {\mathcal O}(\varepsilon^{0}) \\
  \hat{I}_3^{(2)}&=&\Gamma(1+\varepsilon)\left(\frac{1}{\varepsilon^2\,c_5}+
  \frac{1}{\varepsilon\,c_5}\log(c_1\,c_4)\right)
  + {\mathcal O}(\varepsilon^{0}) \\
  \hat{I}_3^{(3)}&=&\Gamma(1+\varepsilon)\frac{1}{\varepsilon\,(c_1-c_5)}
  \log\left(\frac{c_5}{c_1}\right) + {\mathcal O}(\varepsilon^{0})
\end{eqnarray}
while $\hat{I}_3^{(4)}$ is a three-mass triangle checked to be finite.
The partial differential equation system is given by
\begin{eqnarray}
  \frac{\partial\hat{I}_4^{(2)}}{\partial c_1}&=&
  \frac{2\,\Gamma(1+\varepsilon)}{c_1-c_5}
  \log\left(\frac{c_5}{c_1}\right) \nonumber \\
  \frac{\partial\hat{I}_4^{(2)}}{\partial c_3}&=&0 \nonumber \\
  \frac{\partial\hat{I}_4^{(2)}}{\partial c_4}&=&
  \frac{\Gamma(1+\varepsilon)}{c_4} \left(
  \frac{1}{\varepsilon} + 2\log(c_5)
  \right) \nonumber \\
  \frac{\partial\hat{I}_4^{(2)}}{\partial c_5}&=&
  \frac{2\,\Gamma(1+\varepsilon)}{c_5}
  \left( \frac{1}{\varepsilon} +
  \frac{c_1}{c_1-c_5}\log({c_1})-
  \frac{c_5}{c_1-c_5}\log({c_5})+\log({c_4})
  \right)
\end{eqnarray}
with the solution
\begin{eqnarray}\label{sol3}
  \hat{I}_4^{(2)} &=&\Gamma(1+\varepsilon)\left[
  \frac{1}{\varepsilon}(\log(c_4)+ 2\log(c_5))-
  \left(\log\left(1-\frac{c_1}{c_5}\right)
  -\log(c_1-c_5)\right)^2 -
  \right. \nonumber\\
  && 2\log(c_1)\log\left(1-\frac{c_1}{c_5} \right)
  +2\log(c_4c_1-c_4c_5)\log(c_5) \nonumber\\
  && \left. -2\textup{Li}_2\left( \frac{c_1}{c_5} \right) + k_{3}\right]
\end{eqnarray}
The integration constant $k_{3}$ is evaluated in the appendix and its
value is
\begin{equation}\label{c03}
  k_{3}=\frac{3}{2\varepsilon^2} - 8\zeta(2)
\end{equation}
After some manipulation we find
\begin{equation}
  \hat{I}_4^{(2)} =\Gamma(1+\varepsilon)\left[
  \frac{{{\left( c_4 \right) }}^{2\,\varepsilon}}{2\,\varepsilon^2} +
\frac{{{\left( c_5 \right)}}^{2\,\varepsilon}}{\varepsilon^2} - {\log
\left(
  \frac{{c_4}}{{c_5}}\right)}^2 +
  2\,\textup{Li}_2\left(1- \frac{c_1}{c_5}
  \right)  - 4\,\zeta(2) \right]
\end{equation}
Reintroducing the original variables we have
\begin{equation}
  \hat{I}_4^{(2)} =\Gamma(1+\varepsilon)\left[
  \left(\frac{\alpha_4}{M}\right)^{2\,\varepsilon} 
  \frac{1}{2\,\varepsilon^2}
  + \frac{{{\left( M\,\alpha_5 \right) }}^{2\,\varepsilon}}{\varepsilon^2}
- {\log \left(
  \frac{{\alpha_4}}{{M^2\,\alpha_5}}\right)}^2 +
  2\,\textup{Li}_2\left(1- \frac{\alpha_1}{\alpha_5}
  \right)  - 4\,\zeta(2) \right]
\end{equation}

\subsection{The integral $\hat{I}_4^{(3)}$}

Putting $a_3=0$ in Eq.(\ref{ggg}) we eliminate the massive propagator
and obtain the easy (opposite) two mass box \cite{OO} or the one
external massive box if we take respectively $q^2\neq0$ or $q^2= 0$.
These integrals are well-known and are reported also in \cite{BDK}.
Here we just put these integrals in the kinematics specified in section 3.

\subsubsection{$\hat{I}_4^{(3)}$, $q^2\neq 0$}

After putting $a_3=0$ in Eq.(\ref{ggg}) the denominator is given by
\begin{equation} \label{den3}
  \frac{{a_1}\,{a_4}}{{{\alpha }_1}\,{{\alpha }_4}} +
  \frac{{a_2}\,{a_4}}{{{\alpha }_2}\,{{\alpha }_4}} +
  \frac{\lambda \,{a_1}\,{a_5}}
   {{{\alpha }_1}\,{{\alpha }_5}} +
  \frac{{a_2}\,{a_5}}{{{\alpha }_2}\,{{\alpha }_5}}
\end{equation}
Using Eq.(4.44) from the third paper in ref.\cite{BDK} the integral in the
kinematics 
of section 3 reads
\begin{eqnarray}
  \hat{I}_4^{(3)}&=&\frac{2\,\Gamma(1 + \varepsilon)}{1 - \lambda
  }\,\left[ \frac{1}{\varepsilon^2}\left(\left( {\alpha_1}\,{\alpha_4}
  \right)^\varepsilon
  - \left({\alpha_2}\,{\alpha_4} \right)^\varepsilon +
  \left({\alpha_2}\,{\alpha_5} \right)^\varepsilon -
  \left( \frac{{\alpha_1}\,{\alpha_5}}{\lambda } \right)^\varepsilon\right) -
  \frac{1}{2}\log^2 \left(\frac{{\alpha_2}\,{\alpha_5}}{{\alpha_1}\,{\alpha_4}}
  \right)
  \right. \nonumber \\
  &&\left. - \textup{Li}_2 \left( 1 - \frac{{\alpha_1}}{{\alpha_2}} \right) -
  \textup{Li}_2 \left(1 - \frac{{\alpha_5}}{{\alpha_4}} \right) +
  \textup{Li}_2(1 - \lambda ) - \textup{Li}_2 \left(
  1 - \frac{{\alpha_2}\,\lambda }{{\alpha_1}} \right) -
  \textup{Li}_2 \left(1 - \frac{{\alpha_4}\,\lambda }{{\alpha_5}} \right) \right]
\end{eqnarray}

\subsubsection{$\hat{I}_4^{(3)}$, $q^2= 0$}

Putting $\lambda=0$ in Eq.(\ref{den3}) the denominator of this integrals is given by
\begin{equation}
  \frac{{a_1}\,{a_4}}{{{\alpha }_1}\,{{\alpha }_4}} +
  \frac{{a_2}\,{a_4}}{{{\alpha }_2}\,{{\alpha }_4}} +
  \frac{{a_2}\,{a_5}}{{{\alpha }_2}\,{{\alpha }_5}}
\end{equation}
Using Eqs.(4.27, 4.40) from the third paper in ref.\cite{BDK} the integral
in the kinematics of section 3 reads
\begin{eqnarray}
\hat{I}_4^{(3)}&=&\Gamma(1 + \varepsilon)\,\left[
    \frac{2}{\varepsilon^2}\,\left( {\left( {\alpha_1}\,{\alpha_4} \right)
    }^\varepsilon -
    {\left( {\alpha_2}\,{\alpha_4} \right) }^\varepsilon +
    {\left( {\alpha_2}\,{\alpha_5} \right) }^\varepsilon \right) -
    {\log^2 \left(\frac{{\alpha_1}\,{\alpha_4}}{{\alpha_2}\,{\alpha_5}}\right)} 
    \right. \nonumber \\
    && \left.
    - 2\,\textup{Li}_2\left(1 - \frac{{\alpha_1}}{{\alpha_2}} \right) -
    2\,\textup{Li}_2\left(1 - \frac{{\alpha_5}}{{\alpha_4}}\right)
    -4\,\zeta(2) \right]
\end{eqnarray}

\section{Conclusions}

An expression for the scalar pentagon integral shown in figure 1 can be
built via Eqs.(\ref{ppp}), (\ref{rel}), the four points integrals
evaluated in the last section and the coefficients in table 1. The expressions
for $\hat{I}_5$ are very long and are not reported.

More familiar kinematics is realized by replacing the variables $\alpha_i$,
$\lambda$ and $M^2$ with their definitions in terms of $s_{ij}$, $q^2$
and $m^2$. Tensor integrals will be
considered in a separate paper \cite{INP}, however the simplest one of
them, the vector integral, is related to the scalar integrals
with one Feynman parameter in the numerator by the following relation
\begin{equation}\label{vec}
  I_n^D[l^\mu]\rightarrow I_n^D[\mathcal{P}^\mu]
\end{equation}
in which the arrow means integration over loop momentum $l$, the
integrand numerator is in the square brackets and
\begin{equation}
  {\mathcal P}^\mu=\sum_{i=1}^{n-1}a_{i+1}p_i^\mu
\end{equation}
with $p_i$ given in Eqs.(\ref{pi}). The integrals $\hat{I}_5[a_j]$ can
be evaluated by \cite{BDK,CGM}
\begin{equation}
  \hat{I}_5[a_j]=\frac{1}{2N_5}\sum_{i=1}^5\eta_{ji}\hat{I}_{4}^{(i)}+
  {\mathcal O}(\varepsilon)
\end{equation}
where $\eta$ defined in Eq.(\ref{fff}) is deduced by $\rho$ given in
Eq.(\ref{eee}) {\scriptsize\[
  \eta=\pmatrix{1&-1&1 - \lambda &1 - 2M^2\left( -1 + \lambda  \right) &
  -1 + 2M^2\left( -1 + \lambda  \right)\cr
  -1&1&-1 + \lambda &1 + 2\left( -1 + M^2\left( -1 + \lambda  \right)
  \right)\lambda &1 - 2M^2\left( -1 + \lambda  \right)\cr
  1 - \lambda &-1 + \lambda &{\left( -1 + \lambda  \right) }^2&-1 + \lambda &
  1 - \lambda \cr
  1 - 2M^2\left( -1 + \lambda  \right) &
  1 + 2\left( -1 + M^2\left( -1 + \lambda  \right)  \right)\lambda &
  -1 + \lambda &1&-1\cr
  -1 + 2M^2\left( -1 + \lambda  \right) &
  1 - 2M^2\left( -1 + \lambda  \right) &1 - \lambda &-1&1}\]}

\noindent
Higher tensor integrals can be evaluated considering that they are linked to
scalar integrals with more powers of Feynman parameters in the numerator
\cite{BDK}. Such a decompositioncan can also be organized in a way that
drastically reduces numerical instabilities genarated by the presence of 
inverse powers of Gram determinants \cite{CGM}.
Besides the deep inelastic case mentioned in the introduction, the results
obtained in the present paper with $q^2\neq 0$ can be useful in the
evaluation at one loop of the decay amplitude of a real $W$ boson or a
virtual photon in a heavy
quark-antiquark pair and two light quarks. Let us consider the case in
which all massless particles and $k_5$ are gluons, then the pentagon
studied with $q^2= 0$ can be identified with one of the four pentagon in
the perturbative evaluation of the one-loop associated production of heavy
quark in the gluon-gluon-fusion with a gluon in the final state
($gg\rightarrow Q\overline{Q}g$); in this case, indeed, pentagons are
found in which the propagators form chains with $1$,
$2$, $3$ and $4$ equal mass fermions the first of which is calculated in
the present paper while the other ones can be calculated analogously.

\vspace{1 cm}

The author gratefully acknowledges Prof. P. Strolin who supported the
present research, Profs. G. Cosenza and A. Della Selva for discussions,
Dr. G. De Lellis for suggesting the topic and for many comments on
the manuscript, Dr. D. Falcone for suggestions on the manuscript, Drs. 
R. Mertig and F. Orellana for help with $FeynCalc$ \cite{FC}, Dr. F.
Di Capua and Dr. L. Scotto Lavina for a quick help with x-fig and Dr. G.
Celentano for help with LaTeX.

\appendix
\section{Four points initial conditions}

In this appendix we report the calculation of the integration
constants for the four points integral of section four systematically
neglecting ${\mathcal O}(\varepsilon)$ terms. Instead of reporting all
length passages, we give the steps that can be followed by programs of
function manipulation like the used Mathematica.

\subsection{Integration constant for $\hat{I}_4^{(1)}$ integral}

The point chosen to evaluate $\hat{I}_4^{(1)}$ is
$c_2=c_3=(c_4/2)=c_5=1$ where the expression
in Eq.(\ref{sol1}) gives
\begin{equation}\label{ap1}
  \hat{I}_4^{(1)}=\Gamma(1+\varepsilon)\left(\frac{1}{\varepsilon}\,\log(2)-
  3\,\zeta(2) +k_{1}
  \right).
\end{equation}
The expression for the integral at the point selected deduced using
Eq.(\ref{quat}) and the first of Eqs.(\ref{fff}) is
\begin{equation}
  \hat{I}_4^{(1)}=\frac{\Gamma(2+\varepsilon)}{2}\int_0^1 d^4a_i
  \,\frac{\delta(1-\sum_ia_i)}{(a_3^2+\frac{1}{2}\,a_2a_4
  +a_2a_5+a_3a_5)^{2+\varepsilon}}.
\end{equation}
The factor $2$ is given by $\Pi c_i$. Renaming $a_2$ with $x$,
$a_3$ with $y$ and $a_4$ with $z$, and performing the transformation
$x\rightarrow 1-x$, $y\rightarrow x-y$ and $z\rightarrow z$ we arrive
at the expression
\begin{equation}
  \hat{I}_4^{(1)}=\frac{\Gamma(2+\varepsilon)}{2}\int_0^1 dx \int_0^x dy
  \int_0^y dz \,\frac{1}{(x^2-2\,x\,y-\frac{1}{2}\, x\, z + y \,z+y-
  \frac{1}{2}\,z)^{2+\varepsilon}}.
\end{equation}
Putting apart the Gamma function for the moment, the $z$ integration
gives
\begin{equation}
  \frac{-{\left( x^2 + y - 2\,x\,y \right) }^{-1 - \varepsilon}}
   {\left( 1 + \varepsilon \right) \,\left( 1 + x - 2\,y \right) } +
  \frac{2^{1 + \varepsilon}\,{\left( 2\,x^2 + y - 5\,x\,y +
  2\,y^2 \right) }^{-1 - \varepsilon}}
   {\left( 1 + \varepsilon \right) \,\left( 1 + x - 2\,y \right) }.
\end{equation}
The two integrals can be evaluated by shifting both in $y$
\begin{equation}
  y\rightarrow y+\frac{x}{2}
\end{equation}
simplifying the $x$ integral
\begin{equation}
  - \frac{2^{1 + \varepsilon}\,{\left( x + 2\,y - 4\,x\,y \right) }^{-1 - \varepsilon}}
     {\left( 1 + \varepsilon \right) \,\left( 1 - 2\,y \right) } +
  \frac{2^{2 + 2\,\varepsilon}\,{\left( x + 2\,y - 6\,x\,y + 4\,y^2 \right) }^{-1 - \varepsilon}}
   {\left( 1 + \varepsilon \right) \,\left( 1 - 2\,y \right) }
\end{equation}
and inverting the integration order \cite{tHV}. After some
manipulation and expanding some Hypergeometric and Generalized
Hypergeometric functions the result is
\begin{eqnarray} \label{n1}
  \hat{I}_4^{(1)} &=& \Gamma(2+\varepsilon) \left( 
  \frac{1}{2 \,\varepsilon^2} - \frac{1}{2 \,\varepsilon} + \frac{1}{\varepsilon}\,\log (2)
  + \frac{1}{2} + 2\,\zeta(2) - \log (2)  \right).
\end{eqnarray}
Finally, making the substitution $\Gamma(2+\varepsilon )=(1+\varepsilon )
\,\Gamma(1 + \varepsilon)$ in Eq.(\ref{n1}) and taking into account
Eq.(\ref{ap1}) we find $k_{1}$ in Eq.(\ref{c01}).

\subsection{Integration constant for $\hat{I}_4^{(2)}$ integral with $q^2 \neq 0$}

To evaluate the integration constant we evaluate the integral in the
point:
\begin{eqnarray}
c_1&=&\frac{1}{\delta -1} \nonumber \\
c_2&=&\frac{1}{\delta-1}\nonumber \\ c_3&=&\frac{1}{1-\delta
}\nonumber \\ c_4&=&-1 \\
\end{eqnarray}
with
\begin{equation}
\Pi c_i=-\frac{1}{(1-\delta)^3}
\end{equation}
The integrations are trivial but the expression is very long. The
$c_i$ chosen cannot be simultaneously positive so we checked the
result in the point $(\delta/2)=c_1=c_3=c_4=c_5=1$ where the
expression in Eq.(\ref{sol2}) gives
\begin{equation} \label{ss2}
  \hat{I}_4^{(2)}=\Gamma(2+\varepsilon)\left( \frac{1}{2\varepsilon^2}
  - \frac{1}{2\varepsilon} + \frac{1}{\varepsilon}\log(2) + 
  \frac{1}{2} + \frac{3}{2}\zeta(2) - \log (2) -
  \frac{1}{2}\log^2(2) \right)
\end{equation}
The expression for the integral at the point selected is
\begin{equation} 
  \hat{I}_4^{(2)}=\Gamma(2+\varepsilon)\int_0^1 d^4a_i
  \frac{\delta(1-\sum_ia_i)}{(a_3^2+a_1a_3
  +a_1a_4+2a_1a_5+a_3a_5)^{2+\varepsilon}}
\end{equation}
Renaming $a_1$ with $x$, $a_3$ with $y$ and $a_4$ with $z$, and
performing the transformation $x\rightarrow 1-x$, $y\rightarrow x-y$
and $z\rightarrow z$ we arrive at the expression
\begin{equation}
  \hat{I}_4^{(2)}=\Gamma(2+\varepsilon)\int_0^1 dx \int_0^x dy
  \int_0^y dz \frac{1}{(-2xy + y z+x+y-z)^{2+\varepsilon}}
\end{equation}
Putting apart the Gamma function for the moment, the $z$ integration
gives
\begin{equation}
  - \frac{{\left( x + y - 2\,x\,y \right) }^{-1 - \varepsilon}}
  {\left( 1 + \varepsilon \right) \,\left( 1 - y \right) }  +
  \frac{{\left( x - 2\,x\,y + y^2 \right) }^{-1 - \varepsilon}}{\left( 1 +
  \varepsilon \right) \,\left( 1 - y \right) }
\end{equation}
Performing the $x$ integration before and adding and subtracting terms
we find Eq.(\ref{ss2}).

\subsection{Integration constant for $\hat{I}_4^{(2)}$ integral with $q^2= 0$}

The point chosen to evaluate $\hat{I}_4^{(2)}$ is given by
$2c_1=(c_3/2)=(c_4/2)=c_5=1$ in which the expression in Eq.(\ref{sol3})
gives
\begin{equation}\label{ap3}
  \hat{I}_4^{(2)}=\Gamma(1+\varepsilon)\left(\frac{1}{\varepsilon}\log(2)
  - \log^2(2) + 5\,\zeta(2) + k_{3}.
  \right)
\end{equation}
The expression for the integral at the point selected is
\begin{equation}
  \hat{I}_4^{(2)}=\frac{\Gamma(2+\varepsilon)}{2}\int_0^1 d^4a_i
  \,\frac{\delta(1-\sum_ia_i)}{\left(\frac{1}{4}\,a_3^2+a_1a_3
  +a_1a_4+\frac{1}{2}\,a_3a_5 \right)^{2+\varepsilon}}.
\end{equation}
Renaming $a_3$ with $x$, $a_4$ with $y$ and $a_5$ with $z$, and
performing the transformation $x\rightarrow 1-x$, $y\rightarrow x-y$
and $z\rightarrow z$ we arrive at the expression
\begin{equation}
  \hat{I}_4^{(2)}=\frac{\Gamma(2+\varepsilon)}{2}\int_0^1 dx \int_0^x dy
  \int_0^y dz \,\frac{1}{\left( \frac{1}{4}x^2 - y^2 -\frac{1}{2}x z + y z
  -\frac{1}{2}x + y -\frac{1}{2}z + \frac{1}{4}
  \right)^{2+\varepsilon}}.
\end{equation}
Putting apart the Gamma function for the moment, the $z$ integration
gives
\begin{equation}
  \frac{2^{2 + 2\,\varepsilon}\,{\left( 1 - x \right) }^{-1 - \varepsilon}
  \,{\left( 1 - x + 2\,y \right) }^{-1 - \varepsilon}}
   {\left( 1 + \varepsilon \right) \,\left( 1 + x - 2\,y \right) } -
  \frac{2^{2 + 2\,\varepsilon}\,{\left( 1 - 2\,x + x^2 + 4\,y - 
  4\,y^2 \right) }^{-1 - \varepsilon}}
  {\left( 1 + \varepsilon \right) \,\left( 1 + x - 2\,y \right) }.
\end{equation}
Performing the shift
\begin{equation}
  y\rightarrow y+\frac{x}{2}
\end{equation}
gives
\begin{equation}
  \frac{2^{2 + 2\,\varepsilon}\,{\left( 1 - x \right) }^{-1 - \varepsilon}
  \,{\left( 1 + 2\,y \right) }^{-1 - \varepsilon}}
   {\left( 1 + \varepsilon \right) \,\left( 1 - 2\,y \right) } -
  \frac{2^{2 + 2\,\varepsilon}\,{\left( 1 + 4\,y - 4\,x\,y - 
  4\,y^2 \right) }^{-1 - \varepsilon}}
  {\left( 1 + \varepsilon \right) \,\left( 1 - 2\,y \right) },
\end{equation}
finally performing the $x$ integration before and adding and
subtracting terms before expanding in $\varepsilon$ the result is
\begin{equation} \label{n3}
  \hat{I}_4^{(2)}=\Gamma(2+\varepsilon)\left( 
  \frac{3}{2 \,\varepsilon^2} - \frac{3}{2 \,\varepsilon} +
  \frac{1}{\varepsilon}\log(2) + \frac{3}{2}  - 3\,\zeta(2) - \log (2)- \log^2(2) \right).
\end{equation}
Substituting $\Gamma(2+\varepsilon )=(1+\varepsilon )
\,\Gamma(1+\varepsilon)$ in Eq.(\ref{n3}) and taking into account
Eq.(\ref{ap3}) we find $k_{3}$
in Eq.(\ref{c03}).

\end{document}